\documentclass[aps,prd,twocolumn,superscriptaddress,showpacs,showkeys,a4paper]{revtex4-1}

\usepackage[dvips]{graphicx}
\usepackage{color}
\usepackage{hyperref}
\usepackage{ulem}

\graphicspath{{./figures/}}

\begin{document}

\title{
Search for $n-\bar n$ oscillation in Super-Kamiokande
}

\date{\today}

\newcommand{\AFFicrr}{\affiliation{Kamioka Observatory, Institute for Cosmic Ray Research, University of Tokyo, Kamioka, Gifu 506-1205, Japan}}
\newcommand{\AFFkashiwa}{\affiliation{Research Center for Cosmic Neutrinos, Institute for Cosmic Ray Research, University of Tokyo, Kashiwa, Chiba 277-8582, Japan}}
\newcommand{\AFFbu}{\affiliation{Department of Physics, Boston University, Boston, MA 02215, USA}}
\newcommand{\AFFbnl}{\affiliation{Physics Department, Brookhaven National Laboratory, Upton, NY 11973, USA}}
\newcommand{\AFFucd}{\affiliation{Department of Physics, University of California, Davis, Davis, CA 95616, USA}}
\newcommand{\AFFuci}{\affiliation{Department of Physics and Astronomy, University of California, Irvine, Irvine, CA 92697-4575, USA }}
\newcommand{\AFFcsu}{\affiliation{Department of Physics, California State University, Dominguez Hills, Carson, CA 90747, USA}}
\newcommand{\AFFcnm}{\affiliation{Department of Physics, Chonnam National University, Kwangju 500-757, Korea}}
\newcommand{\AFFduke}{\affiliation{Department of Physics, Duke University, Durham NC 27708, USA}}
\newcommand{\AFFgmu}{\affiliation{Department of Physics, George Mason University, Fairfax, VA 22030, USA }}
\newcommand{\AFFgifu}{\affiliation{Department of Physics, Gifu University, Gifu, Gifu 501-1193, Japan}}
\newcommand{\AFFuh}{\affiliation{Department of Physics and Astronomy, University of Hawaii, Honolulu, HI 96822, USA}}
\newcommand{\AFFui}{\affiliation{Department of Physics, Indiana University, Bloomington,  IN 47405-7105, USA}}
\newcommand{\AFFkanagawa}{\affiliation{Physics Division, Department of Engineering, Kanagawa University, Kanagawa, Yokohama 221-8686, Japan}}
\newcommand{\AFFkek}{\affiliation{High Energy Accelerator Research Organization (KEK), Tsukuba, Ibaraki 305-0801, Japan }}
\newcommand{\AFFkobe}{\affiliation{Department of Physics, Kobe University, Kobe, Hyogo 657-8501, Japan}}
\newcommand{\AFFkyoto}{\affiliation{Department of Physics, Kyoto University, Kyoto, Kyoto 606-8502, Japan}}
\newcommand{\AFFlanl}{\affiliation{Physics Division, P-23, Los Alamos National Laboratory, Los Alamos, NM 87544, USA}}
\newcommand{\AFFlsu}{\affiliation{Department of Physics and Astronomy, Louisiana State University, Baton Rouge, LA 70803, USA}}
\newcommand{\AFFumd}{\affiliation{Department of Physics, University of Maryland, College Park, MD 20742, USA }}
\newcommand{\AFFum}{\affiliation{Department of Physics, University of Minnesota, Duluth, MN 55812-2496, USA}}
\newcommand{\AFFmit}{\affiliation{Department of Physics, Massachusetts Institute of Technology, Cambridge, MA 02139, USA}}
\newcommand{\AFFmiyagi}{\affiliation{Department of Physics, Miyagi University of Education, Sendai, Miyagi 980-0845, Japan}}
\newcommand{\AFFnagoya}{\affiliation{Solar Terrestrial Environment Laboratory, Nagoya University, Nagoya, Aichi 464-8602, Japan}}
\newcommand{\AFFsuny}{\affiliation{Department of Physics and Astronomy, State University of New York, Stony Brook, NY 11794-3800, USA}}
\newcommand{\AFFniigata}{\affiliation{Department of Physics, Niigata University, Niigata, Niigata 950-2181, Japan }}
\newcommand{\AFFokayama}{\affiliation{Department of Physics, Okayama University, Okayama, Okayama 700-8530, Japan }}
\newcommand{\AFFosaka}{\affiliation{Department of Physics, Osaka University, Toyonaka, Osaka 560-0043, Japan}}
\newcommand{\AFFseoul}{\affiliation{Department of Physics, Seoul National University, Seoul 151-742, Korea}}
\newcommand{\AFFshizuokasc}{\affiliation{Department of Informatics in Social Welfare, Shizuoka University of Welfare, Yaizu, Shizuoka, 425-8611, Japan}}
\newcommand{\AFFshizuoka}{\affiliation{Department of Systems Engineering, Shizuoka University, Hamamatsu, Shizuoka 432-8561, Japan}}
\newcommand{\AFFskk}{\affiliation{Department of Physics, Sungkyunkwan University, Suwon 440-746, Korea}}
\newcommand{\AFFtohoku}{\affiliation{Research Center for Neutrino Science, Tohoku University, Sendai, Miyagi 980-8578, Japan}}
\newcommand{\AFFtokyo}{\affiliation{The University of Tokyo, Bunkyo, Tokyo 113-0033, Japan }}
\newcommand{\AFFtokai}{\affiliation{Department of Physics, Tokai University, Hiratsuka, Kanagawa 259-1292, Japan}}
\newcommand{\AFFtit}{\affiliation{Department of Physics, Tokyo Institute for Technology, Meguro, Tokyo 152-8551, Japan }}
\newcommand{\AFFtsinghua}{\affiliation{Department of Engineering Physics, Tsinghua University, Beijing, 100084, China}}
\newcommand{\AFFwarsaw}{\affiliation{Institute of Experimental Physics, Warsaw University, 00-681 Warsaw, Poland }}
\newcommand{\AFFuw}{\affiliation{Department of Physics, University of Washington, Seattle, WA 98195-1560, USA}}

\AFFicrr
\AFFkashiwa
\AFFbu
\AFFbnl
\AFFuci
\AFFcsu
\AFFcnm
\AFFduke
\AFFgmu
\AFFgifu
\AFFuh
\AFFui
\AFFkek
\AFFkobe
\AFFkyoto
\AFFlanl
\AFFlsu
\AFFumd
\AFFum
\AFFmiyagi
\AFFnagoya
\AFFsuny
\AFFniigata
\AFFokayama
\AFFosaka
\AFFseoul
\AFFshizuoka
\AFFshizuokasc
\AFFskk
\AFFtohoku
\AFFtokai
\AFFtit
\AFFtokyo
\AFFtsinghua
\AFFwarsaw
\AFFuw
%
\author{K.~Abe}
\AFFicrr
\author{Y.~Hayato}
\author{T.~Iida}
\author{K.~Ishihara} 
\author{J.~Kameda}
\author{Y.~Koshio}
\author{A.~Minamino} 
\author{C.~Mitsuda} 
\author{M.~Miura} 
\AFFicrr
\author{S.~Moriyama} 
\author{M.~Nakahata} 
\AFFicrr
% ...

\collaboration{The Super-Kamiokande Collaboration}
\noaffiliation

\begin{abstract}
A search for neutron-antineutron ($n-\bar{n}$) oscillation 
was undertaken in Super-Kamiokande using 
the 1489 live-day or 
$2.45 \times 10^{34}$ neutron-year exposure data.
This 
process 
violates both baryon and baryon minus lepton numbers 
by an absolute value of two units 
and is predicted by a large class of hypothetical models
where the seesaw mechanism is incorporated to explain the observed 
tiny neutrino masses
and the matter-antimatter asymmetry 
in the Universe.
No evidence for $n-\bar{n}$ oscillation was found, the lower limit of 
the lifetime for neutrons bound in ${}^{16}$O, 
in an analysis that included all of
the significant sources of experimental uncertainties,
was determined to be 
$1.9 \times 10^{32}$~years
at the 90\% confidence level.
The corresponding 
lower limit for 
the oscillation time 
of
free neutrons was calculated to be 
$2.7 \times 10^8$~s
using a theoretical 
value of the nuclear 
suppression factor of 
$0.517 \times 10^{23}$~s$^{-1}$
and its uncertainty.

\end{abstract}



\maketitle

\section{Introduction}
Searches for baryon number ($B$) 
violating
processes are motivated by various 
grand unification theories (GUTs) and by the observation
first put forward by A. Sakharov in 1967 that $B$, $C$, $CP$ violation, and 
nonequilibrium thermodynamics are needed to explain the 
observed matter-antimatter or
baryon number asymmetry of the Universe~\cite{Sakharov}.
However recent experimental limits on  
$p\rightarrow e^{+}\pi^{0}$~\cite{IMBepi0,KAMepi0,SKepi0} 
and $p\rightarrow \bar{\nu}K^{+}$~\cite{Kobayashi:2005pe}  
have
already ruled out the simplest $B-L$ conserving GUTs,
minimal $SU(5)$ and  the minimal supersymmetric version of $SU(5)$,
where $L$ stands for lepton number.
It has also been shown that any $B-L$ conserving baryon
asymmetry generated in the early Universe (above 10 TeV scale)
should be washed out until now
by the triangle anomalies involving electroweak bosons.
Therefore the search for $B-L$ violating reactions
has become increasingly important as a potential explanation 
of the observed baryon number asymmetry in the Universe.

Neutron-antineutron ($n-\bar{n}$) oscillation, a process that violates
$B$ and $B-L$
by two units ($|\Delta B|=2$ and $|\Delta (B-L)|=2$),
was first discussed by V. Kuzmin in 1970~\cite{Kuzumin}.
The discovery of neutrino oscillations~\cite{Fukuda:1998mi}
has renewed interest in theories 
with Majorana spinors which yield $B$ and $L$ symmetry breaking by 
allowing $|\Delta (B-L)|=2$,
with $|\Delta L|$=2 as in neutrinoless
double beta decay and $|\Delta B|$=2 for $n-\bar{n}$ oscillation~\cite{moha,Phillips:2014fgb}.
These models include a large class of supersymmetric and 
right-left symmetric $SU(2)_L \otimes SU(2)_R \otimes SU(4)_C$ theories
that include the seesaw mechanism to generate neutrino masses~\cite{ref3}.
Neutron-antineutron oscillation has also been predicted by GUTs with large
or small extra space-time dimensions~\cite{ref1,ref2}.

An important property of $n-\bar{n}$ oscillation is its dependence on 
a six quark operator, with a mass scaling of 
$M^{-5}$ instead of $M^{-2}$ 
as in the charged X
and Y bosons that mediate nucleon decay in minimal $SU(5)$. Therefore, the 
observation of a significant $n-\bar{n}$ oscillation signal at 
Super-Kamiokande would imply new physics at a scale of approximately 100 TeV, 
a factor of 10 larger than the maximum energies scheduled for study 
at the Large Hadron Collider.

The previous best 90 \% confidence level (C.L.) lifetime limits 
for bound neutrons are from 
IMB with 1.7 and  2.4 $\times 10^{31}$ year~\cite{ref4} in two different 
analyses, Kamiokande with $4.3 \times 10^{31}$ year in oxygen~\cite{ref5}, 
Soudan 2 with $7.2 \times 10^{31}$ year in iron~\cite{ref6}, 
and Frejus with $6.5 \times 10^{31}$ year in iron~\cite{ref9}.
The current best limit on the oscillation time of unbound neutrons is given 
by ILL (Grenoble) as $0.86 \times 10^{8}$ seconds~\cite{ref7}.
The lifetime limit for bound neutrons can be converted to the 
$n-\bar{n}$ oscillation time for free neutrons by the relationship 
\begin{equation}
T_{n-\bar{n}}=R \cdot \tau^2_{n-\bar{n}}
\label{eqn:tandtau}
\end{equation}
where $\tau_{n-\bar{n}}$ and $T_{n-\bar{n}}$ are the oscillation time of 
a free neutron and the lifetime of a bound neutron, and $R$ is the nuclear 
suppression factor, for which theoretical estimates are given in the 
literature~\cite{Friedman_and_Gal}. 

In this paper, we describe the results of our neutron-antineutron oscillation 
search in Super-Kamiokande.
No positive signal was observed, but from our data we obtained a lower limit 
on the lifetime of a neutron in oxygen of
$1.9\times10^{32}$ year
(90\% C.L.), which is about 
four
times larger than the previous highest experimental limits.

\section{The Super-Kamiokande Detector}
Super-Kamiokande is a ring imaging water Cherenkov 
detector
containing 50 ktons of ultrapure water, located in Kamioka-town in Gifu prefecture, Japan. 
The detector is composed of the inner detector, which contains 
22.5 kton fiducial volume, and the outer detector
for tagging cosmic ray muons entering the detector.
Descriptive overviews of Super-Kamiokande and technical 
details are given in the literature~\cite{ref10}.
The $n-\bar{n}$ oscillation analysis described in this paper uses 
the complete Super-Kamiokande-I data set, which is equivalent to 
$2.45 \times 10^{34}$ neutron-year exposure data.
This data set of 1489 live-days
was taken from May 31, 1996 to July 15,
2001.

An antineutron is expected to annihilate quickly
with one of the surrounding nucleons and to produce multiple secondary hadrons, mainly pions.
The total momentum of the secondaries is nearly zero,
except for the Fermi momenta of the annihilating $\bar{n}$ and nucleon, and the
total energy is nearly equal to the total mass of the two nucleons. 
Therefore in the case of no nuclear interactions it was expected that $n-\bar{n}$ oscillation events in 
oxygen
would 
produce mesons with
a total energy of about 2~GeV,
distributed isotropically.  
\section{Monte Carlo simulation}

A detailed Monte Carlo (MC) simulation was formulated for this study.
Since the literature on ($\bar{n}$ + nucleon) annihilation in nuclei is sparse,
$\bar{p}p$ and $\bar{p}d$ data from hydrogen and deuterium bubble 
chambers~\cite{ref11,ref12,ref13}
were used to determine the branching ratios for the annihilation final states. The branching ratios included in our simulations are displayed in Table~\ref{table:branch}.
The values of kinematic quantities in our MC were determined using relativistic phase-space distributions that included the Fermi momentum of the annihilating nucleons.
Pions and omegas produced in the $\bar{n}$-nucleon annihilations were 
propagated through the residual nucleus.
This study used a program originally developed for the IMB experiment 
to simulate meson-nucleon interactions~\cite{ref4}.
The cross sections for the pion-residual nucleus interactions were based on 
an interpolation from measured
pion-carbon and pion-aluminum cross sections to those for pion-$^{16}$O 
interactions.
Excitation of the $\Delta$(1232) resonance was the most important effect 
in the nuclear propagation phase. It was assumed that
the pion and omega cross sections scaled linearly with matter density, 
a quantity that decreases as the mesons move
away from the annihilation point and exit from the nucleus. The Fermi 
momentum of the interacting nucleon and the possibility of Pauli blocking
were also included in our simulations.
We found that 49\% of the pions did not interact, while 24\% were absorbed 
and 3\% interacted with a nucleon to produce an additional pion or occasionally two more pions, and
the rest of the pion interactions involved scattering.
These interaction probabilities yielded total and charged pion multiplicities of 3.5 and 2.2, respectively, with an average charged 
pion momentum of 310 MeV/$c$ and a root mean square width of 190 MeV/$c$. There was a probability of 0.56\% that an event 
included an $\omega^0$ that emerged from the nucleus without having decayed.
The final states of the nuclear fragments were calculated using an algorithm based on a simulation from Oak Ridge National Laboratory~\cite{ORNL}.
Fragments of the residual nucleus that contained two or more nucleons were not simulated in water, since
most of the nucleons in these fragments had momenta that were below the threshold for inelastic hadron interactions.
Therefore only free $n$ and $p$ fragments were propagated through water. 


Since the largest source of background events for $n-\bar{n}$ oscillation is 
atmospheric neutrinos, we prepared a large MC event sample that corresponded 
to an atmospheric neutrino exposure of 500 years.
A detailed description of this simulation is given in 
reference~\cite{superk_fullpaper}.
The propagation of particles and Cherenkov light in the detector was modeled by a 
program that is based on the GEANT-3 package~\cite{GEANT3}. 
The detector geometry, the generation and propagation of Cherenkov radiation from charged particles, and the response of the photomultiplier tubes (PMTs)
and data acquisition electronics were also included in our simulations.
Hadron interactions were simulated using the CALOR package~\cite{CALOR} for nucleons and charged pions with $p_{\pi}> 500$~MeV/$c$,
and through a custom-made program~\cite{kamiokande-sim} for charged pions 
of $p_{\pi} < 500$~MeV/$c$.

\begin{table}[h]
\caption{The branching ratios for the $\bar{n}$+nucleon annihilations in our simulations. These factors were derived from $\bar{p}p$ and $\bar{p}d$ bubble chamber data\cite{ref11,ref12,ref13}.}
\label{table:branch}
\[\begin{tabular} {lrclr}
\hline\hline
\multicolumn{2}{c}{$\bar{n}$+$p$} & ~~~~~~~~~~~~ &\multicolumn{2}{c}{$\bar{n}$+$n$} \\
\hline
$\pi^{+}\pi^{0}$             & 1\%  && $\pi^{+}\pi^{-}$              & 2\%    \\ 
$\pi^{+}2\pi^{0}$            & 8\%  && $2\pi^{0}$                    & 1.5\% \\ 
$\pi^{+}3\pi^{0}$            & 10\% && $\pi^{+}\pi^{-}\pi^{0}$       & 6.5\% \\ 
2$\pi^{+}\pi^{-}\pi^{0}$     & 22\% && $\pi^{+}\pi^{-}2\pi^{0}$      & 11\%   \\ 
2$\pi^{+}\pi^{-}2\pi^{0}$    & 36\% && $\pi^{+}\pi^{-}3\pi^{0}$      & 28\%   \\ 
2$\pi^{+}\pi^{-}2\omega$     & 16\% && $2\pi^{+}2\pi^{-}$            & 7\%    \\ 
3$\pi^{+}2\pi^{-}\pi^{0}$    &  7\% && $2\pi^{+}2\pi^{-}\pi^{0}$     & 24\%   \\ 
                             &      && $\pi^{+}\pi^{-}\omega$        & 10\%   \\ 
                             &      && 2$\pi^{+}2\pi^{-}2\pi^{0}$    & 10\%   \\ 
\hline\hline
\end{tabular}\]
\end{table}
\section{Data reduction and reconstruction}
The trigger
threshold
for our search 
corresponds to 
5.7~MeV of
electromagnetic energy, therefore signal events would trigger with 100\% 
efficiency.
The majority of triggered events in the Super-Kamiokande data are background
due to inherent radioactivity or to 
cosmic ray muons passing in the detector.
Most of the cosmic ray muon events are rejected by a requirement that the number of hit PMTs in the outer detector within 800 ns of the trigger be less than 25.  
We also require that the reconstructed visible energy be greater than 30 MeV to remove the remaining low energy radioactivity.
This constraint is also necessary to maintain the effectiveness of the event reconstruction.
The reduction algorithms are identical to those used for the atmospheric neutrino analyses and nucleon decay searches~\cite{superk_fullpaper}.

Events remaining after the reduction procedure are processed by the full reconstruction program,
which yields an overall event vertex, the number of Cherenkov rings, and a direction,
particle identification determination, and a momentum for each ring~\cite{superk_fullpaper}.
For $n-\bar{n}$ oscillation events the vertex resolution is 26 cm.
Each Cherenkov ring is identified as being either ``showering'' ($e, \gamma$) or ``nonshowering''
($\mu, \pi, p$) based upon its hit pattern and opening angle. The momentum is subsequently determined using the assigned
particle type and the number of collected photoelectrons inside a cone with an opening half-angle of 70 degrees after corrections for 
geometric effects and light attenuation are made. Our choice of cone angle  
completely covers the Cherenkov cone in water, which has an opening half-angle of about 42 degrees. 
In multiple-Cherenkov ring events, we estimate and separate a sample of photoelectrons for each ring using 
an expected Cherenkov light distribution.

\section{Analysis}
To isolate $n-\bar{n}$ candidates we apply additional criteria to the
fully contained (FC) event sample within the inner detector;
this sample
is used for our atmospheric neutrino 
analyses and nucleon decay searches~\cite{superk_fullpaper}. In addition,
we require the following:
\\
(a) The number of Cherenkov rings $>$ 1, 
(b) 700 MeV $<$ Visible energy $<$ 1300 MeV, 
(c) 0 MeV/$c$ $<$ Total momentum $<$ 450 MeV/$c$ and 
(d) 750 MeV/$c^2$ $<$ Invariant mass $<$ 1800 MeV/$c^2$.
The visible energy is defined as the energy of an electron 
that would produce the same total amount of light
in the detector.
The total momentum is 
$P_{\text{tot}}=|\sum_i^{\text{all-rings}} \overrightarrow{p_i} |$, 
where $\overrightarrow{p_i}$ is the
reconstructed momentum vector of the $i$th ring.
The invariant mass is defined to be $M_{\text{tot}}=
\sqrt{E^2_{\text{tot}}-P^2_{\text{tot}}}$, 
while the total energy is defined as 
$E_{\text{tot}}= \sum_i^{\text{all-rings}} \sqrt{{p_i}^2+{m_i}^2}$, 
where $m_i$ is the mass of the $i$th ring assuming
that showering and nonshowering rings are from $\gamma$ rays and 
charged pions, respectively.
The selection criteria given above were optimized to maximize the ratio 
$\epsilon/\sqrt{b}$,
where $\epsilon$ is the signal detection efficiency and $b$ is 
the number of background events.
  
\begin{figure*}[htbp]
\includegraphics[width=3.5in]{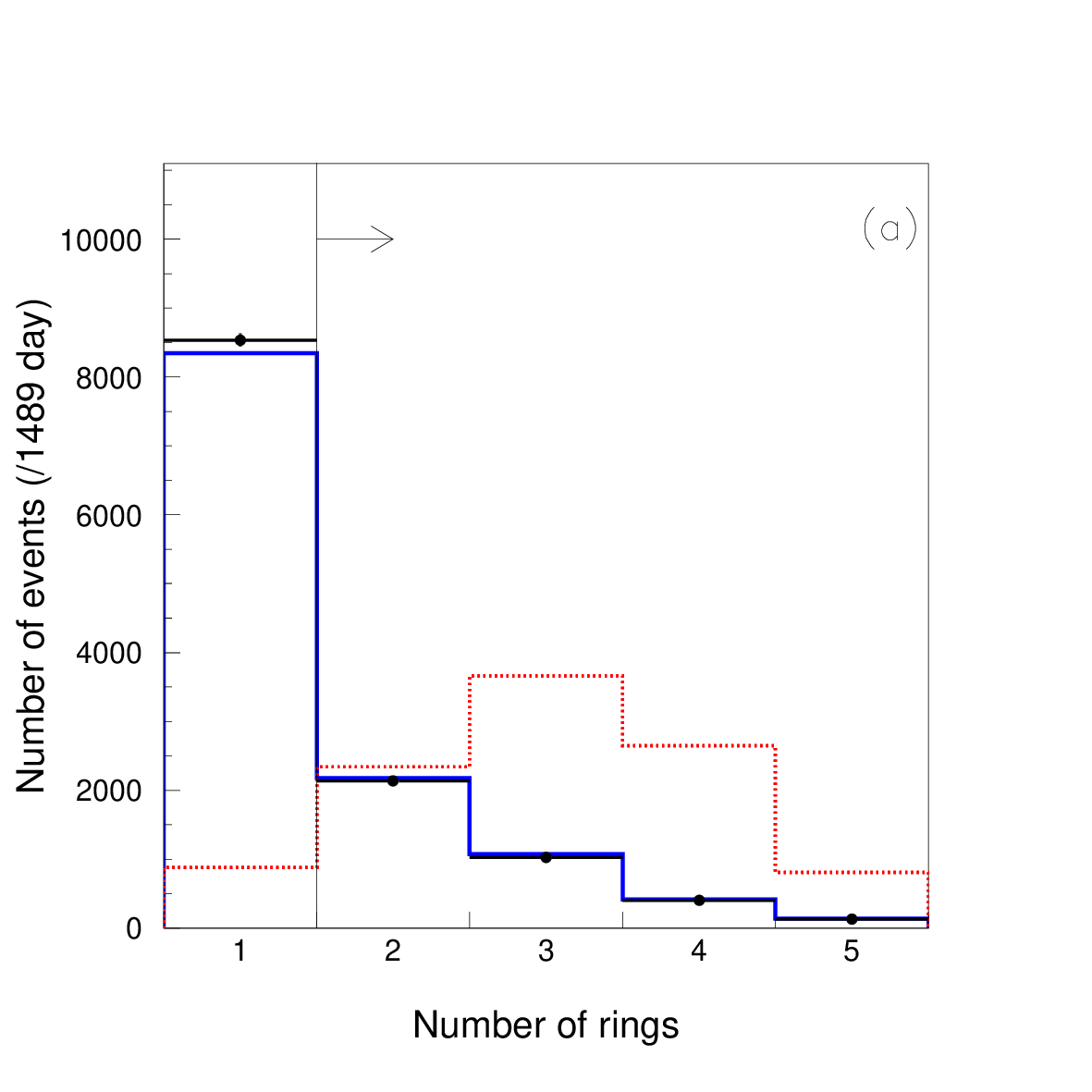}
\includegraphics[width=3.5in]{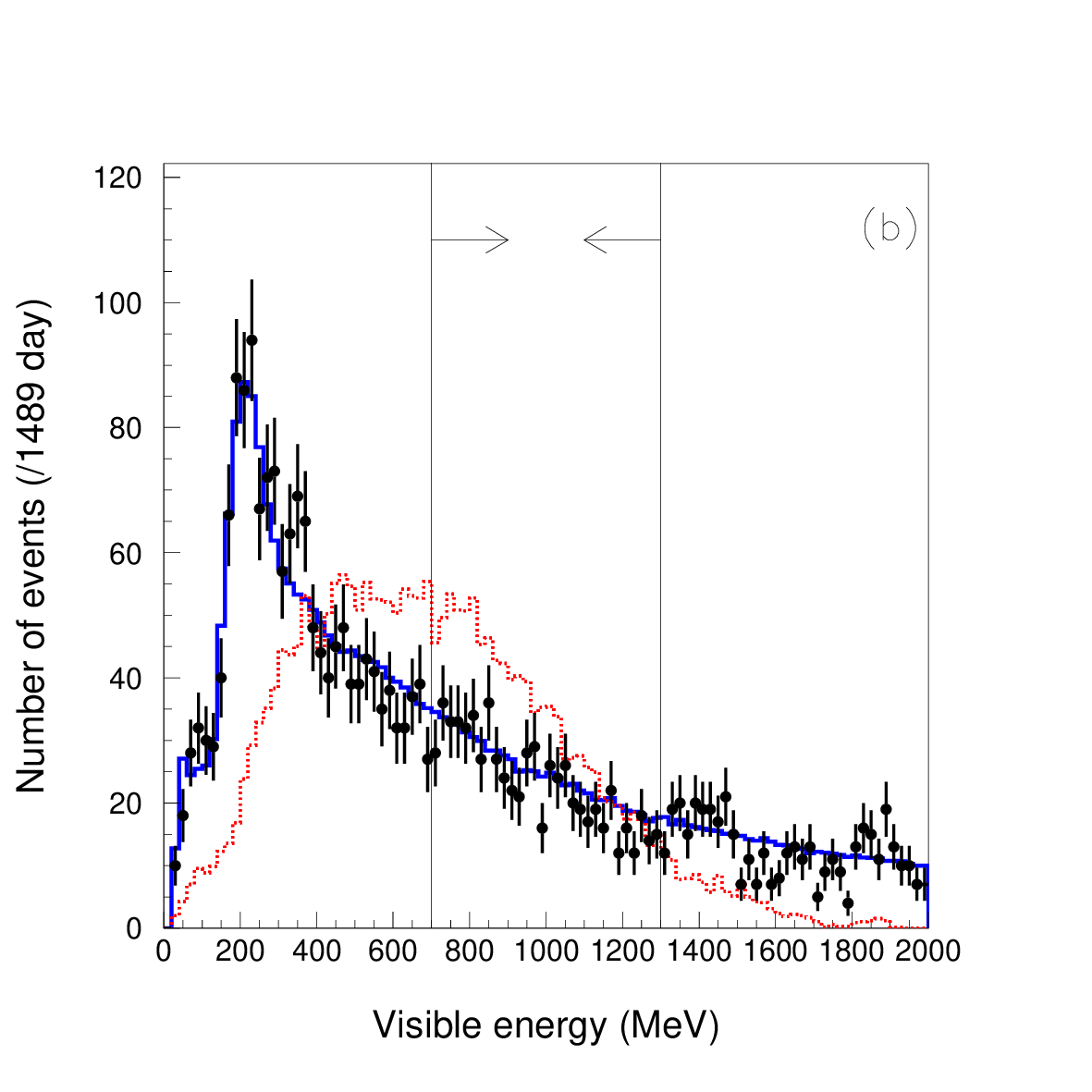}\\
\includegraphics[width=3.5in]{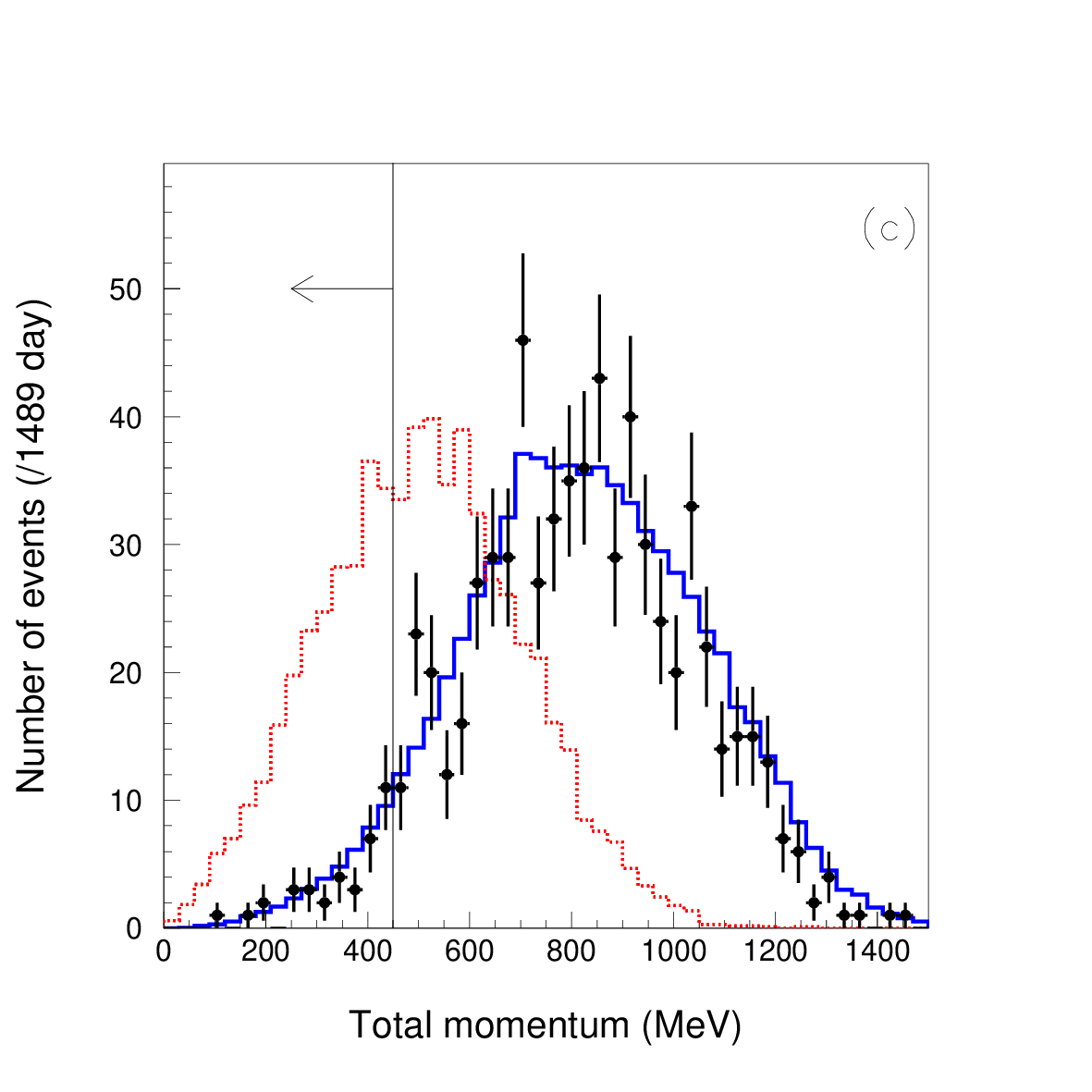}
\includegraphics[width=3.5in]{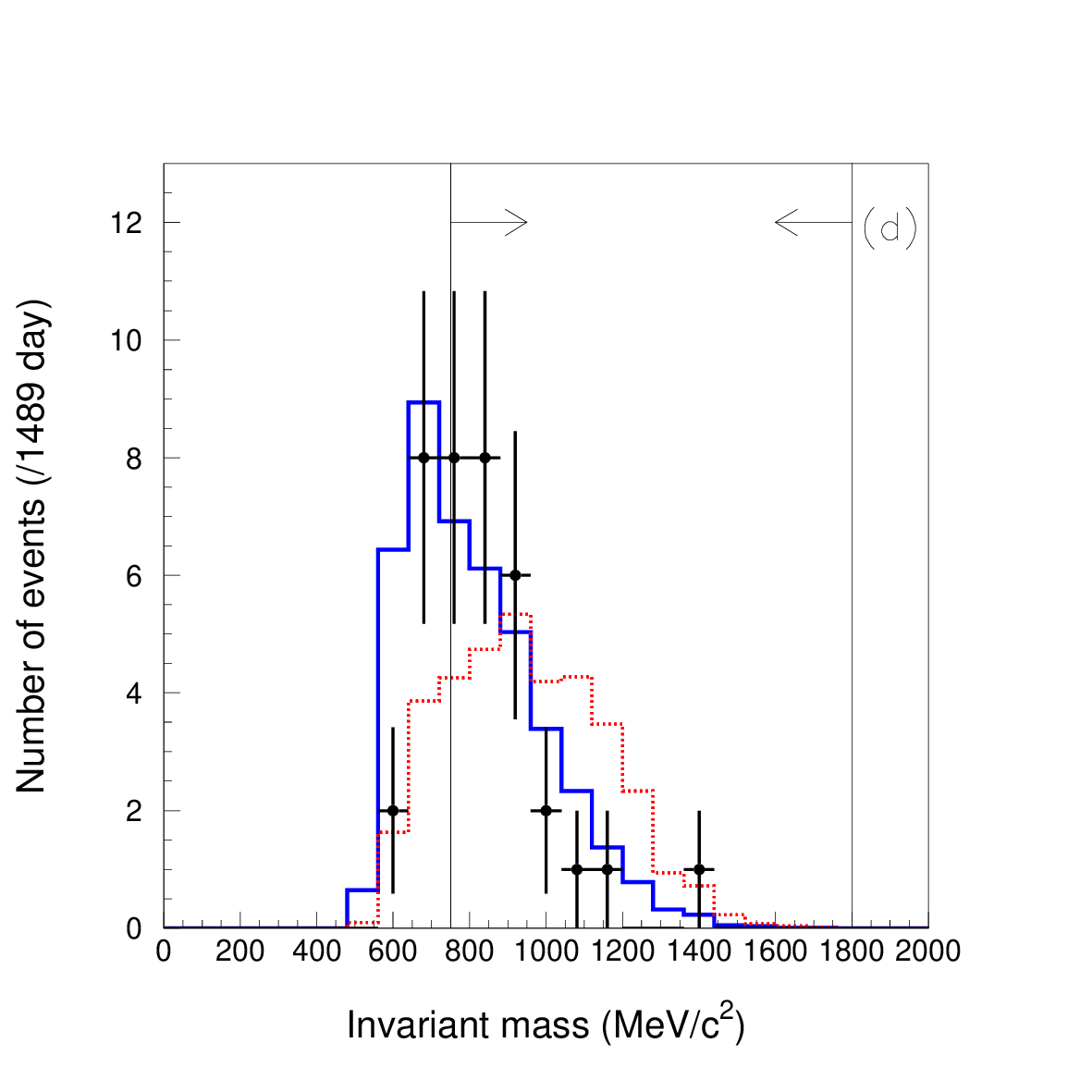}\\
\caption{(color online) The distributions of the kinematic variables 
subject to additional 
criteria at each of the reduction steps described in the text including: 
(a) number of rings, (b) visible energy, (c) total momentum, and
(d) invariant mass. 
Circles indicate data points with statistical error bars; 
solid blue lines and dashed red lines represent the atmospheric neutrino MC 
and the $n-\bar n$ MC, respectively.
Light vertical lines with arrows indicate the event selection criteria.
}
\label{figure:variables}
\end{figure*}

Distributions of the four reconstructed kinematic variables are displayed 
in 
Fig.~\ref{figure:variables}.
Our atmospheric neutrino MC includes the effects of $\nu_{\mu}$ to $\nu_\tau$ neutrino oscillations, 
with mixing parameters of ($\sin^2 2\theta$, $\Delta m^2$)=($1.0$, $2.1 \times 10^{-3}$ eV$^2$) obtained from 
our publication on this topic~\cite{superk_fullpaper}.
Figure~\ref{figure:variables}~(b) shows the visible energy distribution after 
the event selection criterion (a) based on the number of Cherenkov rings 
was imposed.
The average visible energy for $n-\bar{n}$ MC oscillation events is 
about 700 MeV, much 
lower
than twice the nucleon mass. 
The reasons for observing lower visible energy include the mass of charged 
mesons, energy loss by mesons scattering off nucleons, 
and the absorption of mesons in nuclear reactions. 
The visible energy would approach twice the nucleon mass only if 
the annihilation resulted fully in electromagnetic showers 
(such as multiple $\pi^0$ that decayed before interaction).



\begin{figure}[t]
  \includegraphics[width=3.4in]{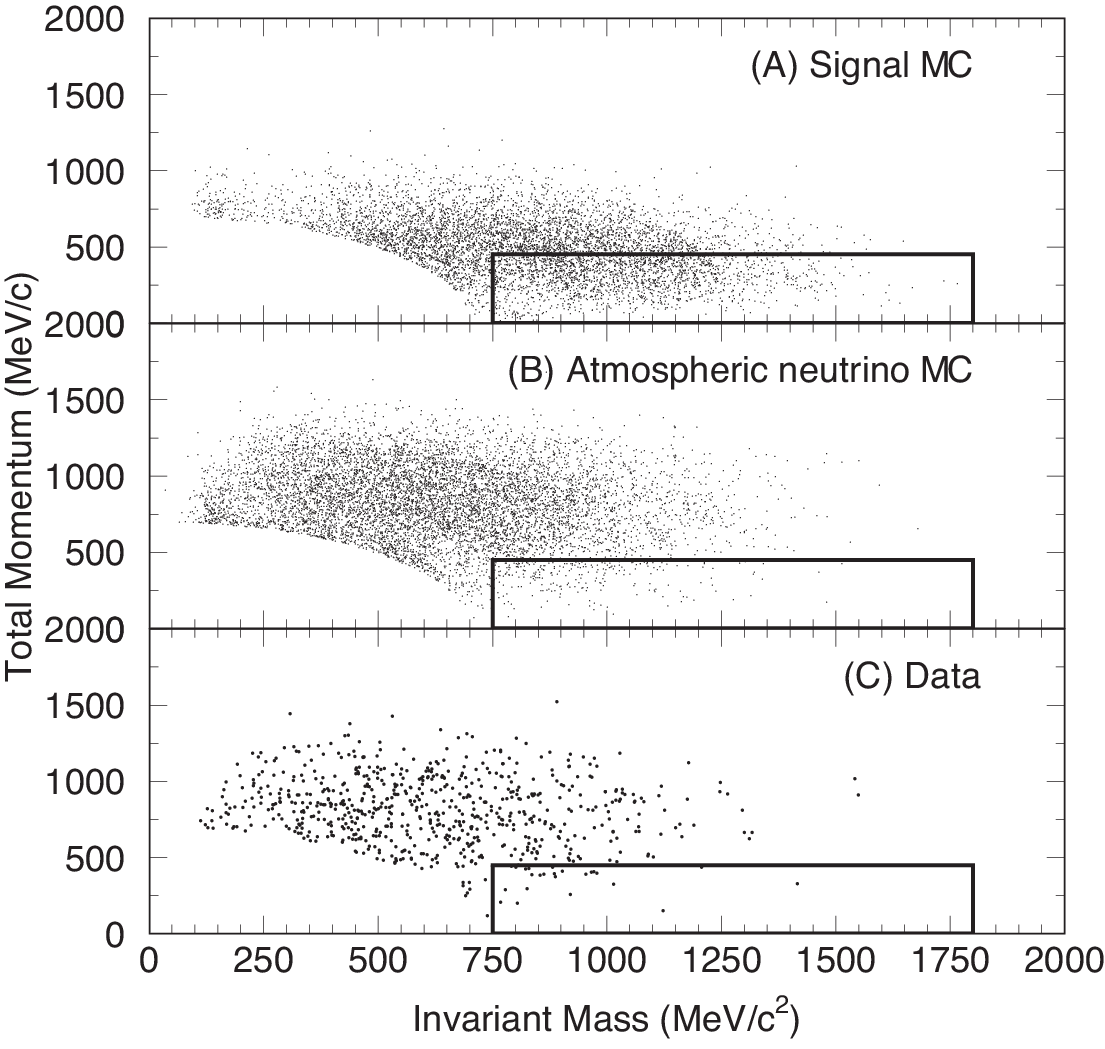}
  \caption{Total momentum vs the invariant mass after applying the 
selection criteria (a)-(b) on the FC sample: 
(A) signal MC, (B) atmospheric neutrino MC, and (C) data.
The boxed region in each panel shows the criterion (c)-(d) 
for the $n-\bar{n}$ oscillation signal.
}
  \label{figure:scatter}
\end{figure}

Application of the event selection criteria (a)-(d) yielded 24 candidate events, a detection efficiency of 12.1\% 
from the remaining $n-\bar{n}$ MC events in the final sample, 
and 24.1 background events as estimated for the 1489 days of 
Super-Kamiokande-I. 
Final event samples are displayed as scattered dots inside of the box-shaped 
scatterplots that apply criteria (c)-(d) in 
Fig.~\ref{figure:scatter}.
Of the remaining background, 57.8\% comes from multihadron production by deep inelastic scattering (DIS),
31.4\% from single pion production via resonances, and 5.5\% from single $\eta$ and $K$ meson production via resonances.
The remaining 5.2\% of the background results from charged current (CC) quasielastic (QE) scattering
and neutral current (NC) elastic scattering accompanied by energetic knocked-out nucleons in the water.
\section{Systematic errors}

Uncertainties in the detection efficiency, exposure, and background rates are given in Table~\ref{table:syserror}.
The major sources of errors in the detection efficiency are the results of uncertainties in the models for propagation of pions and omega mesons through the residual nucleus. In particular, the error due to uncertainties in the $\pi$-nucleon cross section is 20.0\%, as estimated
from the $\pi$-$^{16}$O scattering data shown in~\cite{ref4} and by comparing results from two independent nuclear interaction programs originally developed by IMB ~\cite{ref4} and by Kamiokande~\cite{ref5} (6.1\%).
The uncertainty from the ($\bar{n}$ + nucleon) annihilation branching ratios 
is estimated to be 4.6\% by
comparing different MC results based on variations in the assumed branching 
ratios~\cite{ref24,ref25}.
The 0.6\% asymmetry in the detector gain and the 2.0\% difference 
in the energy scale between the data and the MC
would affect the momentum cut and contribute 1.7\% and 0.4\% uncertainties, 
respectively.
By comparing real data with MC events we estimate that the uncertainty in the Cherenkov ring finding procedure is 2.2\%.
The detection efficiency is estimated to have a total uncertainty of 22.9\% while the exposure uncertainty was found to be 3.0\% through an estimate of the error in the calculation of the effective fiducial volume.

The uncertainties in the atmospheric neutrino flux, the atmospheric neutrino 
energy spectrum, and the  components of the atmospheric neutrino flux 
contribute 7.8\% to the systematic uncertainty of the background rate.
The DIS cross section uncertainty in the low $q^2$ region yields a large contribution to the uncertainty in the background rate, which is estimated to be 14.1\% by comparing the results from two different
parametrizations of the parton distribution 
functions~\cite{GRV94,Bodek}.
The uncertainty in the energy scale, the lack of perfect uniformity 
in the PMT gain, and the uncertainty in the Cherenkov ring finding
contribute a total of 14.9\% to the systematic errors.
The total systematic uncertainty in the background rate is estimated 
to be 23.7\%.

\begin{table} [h]
\caption{Systematic uncertainties in the signal efficiency, exposure, and background rate.}
\label{table:syserror}
\[ \begin{tabular} {llr}

    \multicolumn{3}{c}{Signal efficiency}  \\ \hline\hline
    \multicolumn{2}{c}{Sources} & Uncertainty ($\%$) \\ \hline
    \,&Fermi momentum of nucleons& 6.2 \\
    \,&Branching ratio of $\bar{n}$+nucleons& 4.6 \\
    \,&$\pi$ propagation modeling & 6.1\\
    \,&$\pi$-nucleon cross section in the nucleus & 20.0\\ \hline
    \,&Energy scale & 1.7 \\
    \,&Asymmetry of detector gain & 0.4 \\
    \,&Cherenkov ring finding & 2.2\\
\hline
    \multicolumn{2}{l}{Total} & 22.9\\
\hline\hline
    &&\\
    \multicolumn{3}{c}{Exposure} \\ \hline\hline
    \multicolumn{2}{c}{Sources} & Uncertainty ($\%$) \\ \hline
    \,&Fiducial volume & 3.0\\
    \,&Detector live-days & $<$0.1 \\
\hline
    \multicolumn{2}{l}{Total} & 3.0\\
\hline\hline
  && \\
    \multicolumn{3}{c}{Background rate}                      \\ \hline\hline
    \multicolumn{2}{c}{Sources}         &   Uncertainty($\%$) \\ \hline
   \,& $(\nu_e+\bar{\nu}_e)/(\nu_\mu+\bar{\nu}_\mu)$, $\nu/\bar{\nu}$  ratio   &    0.1, 1.0 \\
   \,& Up/down, horizontal/vertical flux ratios   &     $\ll$1\\
   \,& $K/\pi$ ratio                          &      3.1\\
   \,& Neutrino energy spectrum ($<$1, $>$1GeV) &      6.1,3.6\\ 
\hline
   \,&Neutrino cross sections               &      \\
   \,&\,\,\,QE & 5.8 \\
   \,&\,\,\,1-$\pi$ production &  2.4\\
   \,&\,\,\,DIS & 14.1 \\
   \,&\,\,\,coherent $\pi$ productions & $\ll$1 \\
   \,& CC/NC cross section ratio &  5.0\\
   \,& Axial vector mass in QE and 1$\pi$ prod. & 0.6 \\
   \,& Fermi momentum for QE&   $\ll$1\\
   \,& $\pi$ propagation in $^{16}$O &  4.1\\
 \hline
   \,& Energy scale                  & 4.8 \\
   \,& Asymmetry of detector gain    & 0.5\\
   \,& Cherenkov ring finding        & 14.1\\ \hline
     \multicolumn{2}{l}{Total}        &  23.7\\
\hline\hline
\end{tabular} \]
\end{table}
\section{Results}

No significant excess was found in our full 1489 day Super-Kamiokande-I 
data set.
The lower limit on the lifetime of a neutron bound inside an oxygen nucleus
due to $n-\bar n$ oscillation was calculated from the 24 observed candidate and 24.1 expected background events.
All of the systematic uncertainties were included in the limit calculation by employing the Bayesian statistical method~\cite{ref26} as follows:
\begin{eqnarray}
P(\Gamma|n_{\text{obs}}) &=& A \int\int\int \frac{e^{-(\Gamma \lambda \epsilon +b)}(\Gamma \lambda \epsilon +b)^{n_{\text{obs}}}}{n_{\text{obs}}!} \nonumber \\ 
&&\times P(\Gamma)P(\lambda)P(\epsilon)P(b) d\lambda d\epsilon db .
\label{eq:baysian}
\end{eqnarray}
The above normalization constant A in Eq.~(\ref{eq:baysian}) was determined 
by imposing the constraint
$\int_{0}^{\infty}P(\Gamma|n_{\text{obs}})d\Gamma=1$ where 
$\Gamma$, $\lambda = NT$, 
and $\epsilon$ are the true values of the event rate, exposure, 
and detection efficiency, respectively,
for $n-\bar{n}$ oscillation. 
$N$ and $T$ are the number of neutrons
and the live-days,
$n_{\text{obs}}=24$ is the number of observed candidates,
and $b$ is the true mean number 
of background events. 
$P(\Gamma),P(\lambda),P(\epsilon),P(b)$ in Eq.~(\ref{eq:baysian}) 
are the prior probability density functions, which we assume are Gaussian distributions for $\lambda$, $\epsilon$, and $b$.
$P(\Gamma)$ is a flat distribution for $\Gamma \ge 0$ and 0 for negative $\Gamma$.
Finally, the 90\% C.L. 
limit of the neutron lifetime for $n-\bar{n}$ oscillation 
in oxygen
with the inclusion of systematic uncertainties 
is determined from Eqs.~(\ref{eqn:lifetime})-(\ref{eqn:evratelimit}) 
as follows:
\begin{equation}
T_{n -\bar n}=1/\Gamma_{\text{limit}},
\label{eqn:lifetime}
\end{equation}
\begin{equation}
 \text{C.L.}=\int_{0}^{\Gamma_{\text{limit}}}P(\Gamma|n)d\Gamma
\label{eqn:evratelimit}
\end{equation}
where C.L.=0.9. The calculated result is
\begin{equation}
T_{n-\bar n} > 1.9 \times 10^{32} \mbox{ years.}
\label{eqn:lifetimeresult}
\end{equation}

\begin{table}[h]
\caption{A comparison of the Super-Kamiokande results with those of previous $n-\bar n$ experiments using bound neutrons~\cite{ref4,ref5,ref6,ref9}.
The abbreviation SK, SD2, and KAM stands for Super-Kamiokande,
Soudan 2, and Kamiokande, respectively.
}
\label{table:results}

\[\begin{tabular} {lccccc}
\hline\hline
    Experiment                      & SK    & SD2 & Frejus & KAM   & IMB \\
\hline
    Source of neutrons~~~~              & Oxygen   & Iron  & Iron   & Oxygen & Oxygen\\\hline
    Exposure                        &          &       &        &        &         \\
    ($10^{32}$neutron$\cdot$yr)     & 245    &  21.9 & 5.0    & 3.0    & 3.2    \\\hline
    Efficiency(\%)                  &  12.1    & 18.0  & 30.0   & 33.0   &  50.0     \\\hline
    Candidates                      & 24       & 5     & 0      & 0      &  3    \\\hline
    Backgrounds                     & 24.1     & 4.5   & 2.5(2.1) & 0.9  &  --   \\\hline
    $T_{n-\bar{n}}$ ($10^{32}$yr)   & 1.9     & 0.72  & 0.65   & 0.43   & 0.24  \\\hline
    Suppression factor              &          &       &        &        &      \\
    ($10^{23}$sec$^{-1}$)           & 0.517     & 1.4   & 1.4    & 1.0    & 1.0  \\\hline
    $\tau_{n-\bar{n}}$($10^{8}$sec) & 2.7     & 1.3   & 1.2    & 1.2    & 0.88  \\
\hline\hline
\end{tabular}\]
\end{table}

This lifetime for oscillation of a bound neutron is converted to the 
$n-\bar{n}$ oscillation time of a free neutron using 
Eq.~(\ref{eqn:tandtau}).
The result depends on the choice of the nuclear suppression factor, $R$.
A straightforward application of Eq.~(\ref{eqn:tandtau}), 
using our result in Eq.~(\ref{eqn:lifetimeresult}) and the older suppression
factor $R = 1.0 \times 10^{23}$~s$^{-1}$ yields
a limit on the free neutron oscillation time of 
$> 2.4 \times 10^8$~seconds. 
This may be directly compared with the deduced free neutron limits from 
Kamiokande and IMB, as listed in Table~\ref{table:results},
which used the same nuclear suppression factor. 
Since the time of those experiments, Friedman and Gal have 
published an improved calculation of the nuclear suppression 
factor~\cite{Friedman_and_Gal}. 
Their calculation uses nuclear potentials based on recent 
data~\cite{Friedman_and_Gal2}, 
resulting in a significantly lower value of:
\begin{equation}
R = 0.517 \times 10^{23} {\rm ~s}^{-1},
\label{eqn:newsuppressionfactor}
\end{equation}
with a theoretical uncertainty of 20$-$30\%~\cite{Friedman_and_Gal}.
We note that the earlier value of $R$ is well outside this 
theoretical uncertainty, but simply take the latest estimate and 
its uncertainty at face value for interpreting our result. 
The application of Eqs.~(\ref{eqn:tandtau}) and (\ref{eqn:lifetimeresult})
with this value of $R$ results in a limit on the free neutron oscillation 
time of 
$\tau_{n-\bar n} > 3.4 \times 10^{8} \mbox{ s}$
at the 90\% C.L.
We take account of the theoretical uncertainty of 30\% in $R$ 
by applying the same
Bayesian treatment using Eq.~(\ref{eq:baysian}) 
to the normalization. With this consideration, the
free neutron lifetime is more conservatively limited to be:
\begin{equation}
\tau_{n-\bar n} > 2.7 \times 10^{8} \mbox{ s}
\label{eqn:osctimeresult}
\end{equation}
at the 90\% C.L.
This limit represents the best understanding that may be derived
from our search and can be compared to the direct result on
free neutron oscillation from the
ILL/Grenoble experiment~\cite{ref7},
$\tau_{n-\bar n} > 0.86 \times 10^{8}$ seconds.

\section{Summary}
We have searched for neutron-antineutron oscillation with 
$2.45 \times 10^{34}$ neutron-years of exposure of $^{16}$O 
in the Super-Kamiokande-I experiment. A small number of simple selection 
criteria were employed and the search window in total momentum versus 
invariant mass was optimized for this exposure. Twenty-four events pass 
the selection criteria, but the estimated background was $24.1 \pm 5.7$ 
events, and the distribution of events agrees with the expectation from 
atmospheric neutrino interactions. The signal efficiency was 
$12.1 \pm 2.8$\%. We performed a limit calculation that incorporates 
systematic uncertainties, arriving at a bound lifetime limit of 
$T_{n-\bar n} > 1.9 \times 10^{32}$ years. 
This is more than four times greater 
than the best limit from a water Cherenkov experiment, despite having nearly 
eighty times the exposure because these searches are background limited. 
Using a recent calculation of the nuclear suppression factor, 
the negative results of our search for $n - \bar n$ oscillation in 
$^{16}$O is equivalent to a limit for free $n - \bar n$ oscillation time 
of greater than 
$2.7 \times 10^8$~seconds. 
This result is 
three
times more restrictive than the ILL/Grenoble reactor experiment. 
This places strict limitations on ($\Delta B = 2$, $\Delta L = 0$) 
processes in physics beyond the standard model.

%
%
%
%
\section*{ACKNOWLEDGMENTS}
We gratefully acknowledge the cooperation of the Kamioka Mining 
and Smelting Company.  The Super-Kamiokande experiment was built from, 
and has been operated with, funding by the Japanese Ministry of Education, 
Science, Sports and Culture, and the United States Department of Energy.
We also would like to express our gratitude 
to the following agencies for their 
support of our research;
the U.S. National Science Foundation 
(including Grants No. PHY 0401139 and No. PHY 0901048 to CSUDH),
the National Research Foundation of Korea (Grant No. NRF-2009-353-C00046), 
and the National Natural Science Foundation of China.

 
\end{document}